\documentclass[11pt]{amsart}

\usepackage{amscd}
\usepackage{amsmath}
\usepackage{graphicx}
\usepackage{amsfonts}
\usepackage{amssymb}
\begin{document}

\title[Detecting entanglement by entries]
{Detecting entanglement of states by   entries of their density
matrices}

\author{Xiaofei Qi}
\address[Xiaofei Qi]{
Department of Mathematics, Shanxi University , Taiyuan 030006, P. R.
of China;} \email{qixf1980@126.com}

\author{Jinchuan Hou}
\address{Department of
Mathematics\\
Taiyuan University of Technology\\
 Taiyuan 030024,
  P. R. of China}
\email{jinchuanhou@yahoo.com.cn}


\thanks{This work is partially supported by National Natural Science
Foundation of China (No. 10771157), Research Grant to Returned
Scholars of Shanxi (2007-38) and the Foundation of Shanxi
University.}

\begin{abstract}

For any bipartite systems,   a universal entanglement witness of
rank-4 for pure states is obtained and  a class of finite rank
entanglement witnesses is constructed. In addition,  a method of
detecting entanglement of a state only by entries of its density
matrix with respect to some product basis is obtained.

\end{abstract}
\maketitle

\section{Introduction}

 Let $H$ and $K$ be separable complex Hilbert spaces.
Recall that a quantum state is an operator $\rho\in{\mathcal
B}(H\otimes K)$ which is positive and has trace 1. Denote by
${\mathcal S}(H)$ the set of all states on $H$. If $H$ and $K$ are
finite dimensional, $\rho\in{\mathcal S}(H\otimes K)$ is said to be
separable if $\rho$ can be written as
$$\rho=\sum_{i=1}^k p_i \rho_i\otimes \sigma _i,$$
where $\rho_i$ and $\sigma_i$ are states on $H$ and $K$
respectively, and $p_i$ are positive numbers with $\sum
_{i=1}^kp_i=1$. Otherwise, $\rho$ is said to be inseparable or
entangled (ref. \cite{BZ, NC}). For the case that at least one of
$H$ and $K$ is of infinite dimension,  by  Werner \cite{W},  a state
$\rho$ acting on $H\otimes K$ is called separable if it can be
approximated in the trace norm by the states of the form
$$\sigma=\sum_{i=1}^n p_i \rho_i\otimes \sigma _i,$$
where $\rho_i$ and $\sigma_i$ are states on $H$ and $K$
respectively, and $p_i$ are positive numbers with
$\sum_{i=1}^np_i=1$. Otherwise, $\rho$ is called an entangled state.

Entanglement is a basic physical resource to realize various
quantum information  and quantum communication tasks such as
quantum cryptography, teleportation, dense coding and key
distribution \cite{NC}. It is very important but also difficult to
determine whether or not a state in a composite system is
separable or entangled. It is obvious that every separable state
has a positive partial transpose (the PPT criterion). For $2\times
2$ and $2\times 3$ systems, that is, for the case $\dim H =\dim K
= 2$ or $\dim H = 2,\ \dim K = 3$, a state is separable if and
only if it is a PPT state, that is, has positive partial transpose
(see \cite{Hor, Pe}), but the PPT criterion has no efficiency for
PPT entangled states appearing in the higher dimensional systems.
In \cite{CW}, the realignment criterion for separability in
finite-dimensional systems was found, which says that if
$\rho\in{\mathcal S}(H\otimes K)$ is separable, then the trace
norm of its realignment matrix $\rho^R$ is not greater than 1. The
realignment criterion was generalized to infinite dimensional
system by Guo and Hou in \cite{GH}. A most general approach to
characterize quantum entanglement is based on the notion of
entanglement witnesses (see \cite{Hor}). A self-adjoint operator
$W$ acting on $H\otimes K$ is said to be an entanglement witness
(briefly, EW), if $W$ is not positive and ${\rm Tr}(W\rho)\geq 0$
holds for all separable states $\rho$.  It was shown in \cite{Hor}
that, a state $\rho$ is entangled if and only if it is detected by
some entanglement witness $W$, that is, ${\rm Tr}(W\rho)<0$.
However, constructing entanglement witnesses is a hard task. There
was a considerable effort in constructing and analyzing the
structure of entanglement witnesses for finite and infinite
dimensional systems \cite{B,CK,HQ,JB,TG} (see also \cite{HHH1} for
a review). Recently, Hou and Qi in \cite{HQ}  showed that every
entangled state can be recognized by an entanglement witness $W$
of the form $W=cI+T$ with $I$ the identity operator, $c$ a
nonnegative number and $T$ a finite rank self-adjoint operator and
provided a way how to construct them.

Another important criterion for separability of states is the
positive map criterion \cite[Theorem 2]{Hor}, which claims that a
state $\rho\in{\mathcal S}(H\otimes K)$ with $\dim H\otimes
K<\infty$ is separable if and only if $(\Phi\otimes I)\rho\geq 0$
holds for all positive linear maps $\Phi:{\mathcal
B}(H)\rightarrow{\mathcal B}(K)$.  Hou   \cite{H} generalized the
positive map criterion to the infinite dimensional systems and
obtained the following result.

{\bf  Finite rank elementary operator criterion.} (\cite[Theorem
4.5]{H}) {\it Let $H$, $K$ be complex Hilbert spaces and $\rho$ be a
state acting on $H\otimes K$. Then the following statements are
equivalent.}

(1) {\it $\rho$ is separable;}

(2) {\it $(\Phi\otimes I)\rho \geq 0$ holds for every finite-rank
positive
 elementary operator $\Phi :{\mathcal B}(H)\rightarrow
{\mathcal B}(K)$.}

Recall that a linear map $\Phi :{\mathcal B}(H)\rightarrow {\mathcal
B}(K)$ is an elementary operator if there are operators
$A_1,A_2,\cdots, A_r\in{\mathcal B}(H,K)$ and $B_1,B_2,\cdots,
B_r\in{\mathcal B}(K,H)$ such that $\Phi (X)=\sum_{i=1}^rA_iXB_i$
for all $X\in{\mathcal B}(H)$. It is known that an elementary
operator $\Phi$ is finite rank positive if and only  if  there exist
finite rank operators $C_1,\dots,C_k, D_1,\cdots, D_l\in{\mathcal
B}(H,K)$ such that $(D_1,\cdots, D_l)$ is a contractive local
combination of $(C_1,\cdots, C_k)$ and $\Phi
(X)=\sum_{i=1}^kC_iXC_i^\dag-\sum_{j=1}^lD_jXD_j^\dag$ for all
$X\in{\mathcal B}(H)$ (ref. \cite{H} and the references therein).

Therefore, by the finite rank  elementary operator criterion, a
state $\rho$ on $H\otimes K$ is entangled if and only if there
exists a finite rank positive elementary operator $\Phi:{\mathcal
B}(H)\rightarrow {\mathcal B}(K)$ such that $(\Phi\otimes I)\rho$ is
not positive. Here $\Phi$ must be not completely positive (briefly,
NCP). Thus  it is also important and interesting to find as many as
possible  finite rank positive elementary operators  that are NCP,
and then, to apply them to detect the entanglement of states. In
\cite{QH}, some new finite rank positive elementary operators were
constructed and then applied   to get some new entangled states that
can not be detected by the PPT criterion and the realignment
criterion.

Due to the Choi-Jamio{\l}kowski isomorphism, any EW on finite
dimensional system $H\otimes K$ corresponds to a linear positive
map $\Phi:{\mathcal B}(H)\rightarrow{\mathcal B}(H)$. In fact, for
system $H\otimes K$ of any dimension, if $\Phi:{\mathcal
B}(H)\rightarrow{\mathcal B}(H)$ is a normal positive completely
bounded linear map, and if $\rho_0$ is an entangled state on
$H\otimes K$, then $W=(\Phi\otimes I)\rho_0$ is an entanglement
witness whenever $W$ is not positive (see lemma 2.1). Recall that
a linear map $\Delta : {\mathcal B}(H)\rightarrow{\mathcal B}(K)$
is said to be completely bounded if $\Delta\otimes I$ is bounded;
is said to be normal if it is weakly continuous on bounded sets,
or equivalently, if it is ultra-weakly continuous (i.e., if
$\{A_\alpha\}$ is a bounded net and there is $A\in{\mathcal B}(H)$
such that $\langle x|A_\alpha|y\rangle$ converges to $\langle
x|A|y\rangle$ for any $|x\rangle, |y\rangle\in H$, then $\langle
\phi|\Delta(A_\alpha)|\psi\rangle$ converges to $\langle
\phi|\Delta(A)|\psi\rangle$ for any $|\phi\rangle, |\psi\rangle\in
K$. ref. \cite[pp.59]{Dix}).

The finite rank elementary operator criterion, together with lemma
2.1, gives a way of constructing finite rank entanglement witnesses
from finite rank positive elementary operators for both finite and
infinite dimensional bipartite systems.  In the  present paper, we
construct  a rank-4 entanglement witness $W$ that has some what
``universal" property for pure states in any bipartite systems
$H\otimes K$. We show that, for such a rank-4 entanglement witness
$W$, a pure state $\rho$ is entangled if and only if there exist
unitary operators $U$ on $H$ and $V$ on $K$ such that ${\rm
Tr}((U\otimes V)W(U^\dag\otimes V^\dag)\rho)<0$. In addition, if
$\rho$ is a mixed state such that ${\rm Tr}((U\otimes
V)W(U^\dag\otimes V^\dag)\rho)<0$, then $\rho$ is 1-distillable (see
theorem 2.2). We also construct a class of entanglement witnesses
from the finite rank positive elementary operators obtained in
\cite{QH} (see theorem 3.1).

So far, by our knowledge, there is no  methods of recognizing the
entanglement of a state by merely the entries of its density
matrix. Another interesting result of this paper gives a way of
detecting the entanglement of a state in a bipartite system by
only a part of entries of its density matrix (see theorems 3.2,
3.3). This method is simple, computable and practicable because it
provide a way  to recognize the entanglement of a state by some
suitably chosen entries of its matrix representation with respect
to  some given product basis. As an illustration, some new
examples of entangled states that can be recognized by this way
are proposed, which also provides some new entangled states that
can not be detected by the PPT criterion and the realignment
criterion (see examples 3.4, 3.5).

Recall that a bipartite state $\rho$ is called $n$-distillable, if
and only if maximally entangled bipartite pure states, e.g.
$|\psi\rangle= \frac{1}{2}(|11'\rangle + |22'\rangle)$, can be
created from $n$ identical copies of the state $\rho$ by means of
local operations and classical communication; is called distillable
if it is $n$-distillable for some $n$. It has been shown that all
 entangled pure states are distillable. However it is a
challenge  to give an operational criterion of distillability for
general mixed states \cite{Hor1}. In \cite{Hor2}, it was shown  that
a density matrix $\rho $ is distillable if and only if there are
some projectors $P$, $Q$ that map high dimensional spaces to
two-dimensional ones such that the state $(P\otimes Q)\rho^{\otimes
n}(P\otimes Q)$ is entangled for some $n $ copies.

\section{Universal entanglement witnesses for pure states}

In this section we will  give a simple necessary and sufficient
condition  for separability of pure states in bipartite composite
systems of any dimension.

Before stating the main result in this section, we give a basic
lemma.

{\bf Lemma 2.1.} {\it Let $H$, $K$ be complex Hilbert spaces of
any dimension and let $\Phi:{\mathcal B}(H)\rightarrow{\mathcal
B}(H)$ be a positive normal  completely bounded linear map. Then,
for any entangled state $\rho_0$ on $H\otimes K$, $W=(\Phi\otimes
I)\rho_0$ is an entanglement witness whenever $W$ on $H\otimes K$
is not positive.}

{\bf Proof.} Because $\Phi$ is completely bounded, $W=(\Phi\otimes
I)\rho_0$ is a bounded self-adjoint operator on $H\otimes K$. Note
that ${\mathcal B}(H)={\mathcal T}(H)^*$, where ${\mathcal T}(H)$
denotes the Banach space of all trace class operators on $H$
endowed with the trace norm. Then the normality of $\Phi$ implies
that there exists a bounded linear map $\Delta :{\mathcal
T}(H)\rightarrow{\mathcal T}(H)$ such that $\Phi=\Delta^*$. We
claim that $\Delta$ is also positive. In fact, for any unit vector
$|\phi\rangle\in H$ and any positive operator $A\in{\mathcal
B}(H)$, we have
$${\rm Tr}(A\Delta(|\phi\rangle\langle\phi|))={\rm Tr}(\Phi(A)(|\phi\rangle\langle\phi|))
=\langle\phi|\Phi(A)|\phi\rangle\geq 0.$$ This implies that
$\Delta(|\phi\rangle\langle\phi|)$ is positive for any
$|\phi\rangle$. So, $\Delta$ is a positive linear map.

Now, for any separable state $\rho\in{\mathcal S}(H\otimes K)$, we
have
$${\rm Tr}(W\rho)={\rm Tr}((\Phi\otimes I)\rho_0\cdot \rho)={\rm Tr}(\rho_0\cdot (\Delta\otimes
I)\rho)\geq 0
$$
since $(\Delta\otimes I)\rho\geq 0$. So, if $W$ is not positive,
then it is an entanglement witness.\hfill$\Box$

Since every elementary operator is normal and completely bounded, by
Lemma 2.1, if $\Phi$ is a positive elementary operator and if
$\rho_0$ is an entangled state, then $W=(\Phi\otimes I)\rho_0$ is an
entanglement witness whenever $W$ is not positive. Also note that,
if $W$ is an entanglement witness, then for any positive number $b$,
$bW$ is  an entanglement  witness, too.

Let $W$ be an entanglement witness on $H\otimes K$. We say that $W$
is universal (for all states) if, for any entangled state $\rho$ on
$H\otimes K$, there exist unitary operators $U$ on $H$ and $V$ on
$K$ such that ${\rm Tr}((U\otimes V)W(U^\dag\otimes V^\dag)\rho)<0$;
$W$ is universal for pure states if, for any entangled pure state
$\rho$ on $H\otimes K$, there exist unitary operators $U$ on $H$ and
$V$ on $K$ such that ${\rm Tr}((U\otimes V)W(U^\dag\otimes
V^\dag)\rho)<0$.

The following is the main result of this section, which  gives a
universal entanglement witness  of rank-4 for pure states.
Particularly, we conclude that the separability of   pure states can
be determined by a special class of rank-4 witnesses, and every
1-distillable state can be detected by one of such rank-4
entanglement witnesses. However, we do not know whether or not there
exists a universal entanglement witness for all states.

Let ${\mathcal U}(H)$ (resp. ${\mathcal U}(K)$) be the group of all
unitary operators on $H$ (resp. on $K$).

{\bf Theorem 2.2.} {\it Let $H$ and $K$ be  Hilbert spaces and let
$\{|i\rangle\}_{i=1}^{\dim H\leq \infty }$ and
$\{|j'\rangle\}_{j=1}^{\dim K\leq \infty }$ be any orthonormal bases
of $H$ and $K$, respectively. Let
$$W=|1\rangle|2'\rangle\langle 1|\langle2'|-|1\rangle|1'\rangle\langle 2|\langle2'|
-|2\rangle|2'\rangle\langle 1|\langle1'|+|2\rangle|1'\rangle\langle
2|\langle1'|.\eqno(2.1)$$
 Then $W$ is an entanglement witness  of rank-4. Moreover, the following statements are true.}

(1) {\it If $\rho$ is a pure state, then  $\rho$ is separable if and
only if
$${\rm Tr}((U\otimes V)W(U^\dagger\otimes V^\dagger)\rho)\geq
0\eqno(2.2)$$ hold for all $U\in{\mathcal U}(H)$ and $V\in{\mathcal
U}(K)$. So $W$ is a universal entanglement witness for pure states.}

(2) {\it Let $\rho$ be a state. If there exist $U\in{\mathcal U}(H)$
and $V\in{\mathcal U}(K)$ such that ${\rm Tr}((U\otimes
V)W(U^\dagger\otimes V^\dagger)\rho)< 0$, then $\rho$ is entangled
and 1-distillable.}

{\bf Proof.} We first prove that $W$ is an entanglement witness.
It is obvious that $W$ is not positive.  Define a map $
\Phi:{\mathcal B}(H)\rightarrow {\mathcal B}(H)$ by
$$\begin{array}{rl} \Phi
(A)=&E_{11}AE_{11}^\dagger+E_{22}AE_{22}^\dagger+E_{12}AE_{12}^\dagger\\&+E_{21}AE_{21}^\dagger
-(E_{11}+E_{22})A(E_{11}+E_{22})^\dagger
\end{array}\eqno(2.3)
$$
for every $A\in{\mathcal B}(H)$, where $E_{ij}=|i\rangle\langle
j|\in{\mathcal B}(H)$. It is obvious that $\Phi$ is a positive map
because the map
$$\left(\begin{array}{cc} a_{11}&a_{12}\\ a_{21}&
a_{22}\end{array}\right)\mapsto \left(\begin{array}{cc} a_{22}
&-a_{12}\\ -a_{21}& a_{11}\end{array}\right)$$ on $M_2({\mathbb C})$
is positive. Note that  $W=2(\Phi\otimes I)\rho_+$, where
$\rho_+=|\psi_+\rangle\langle\psi_+|$ with
$|\psi_+\rangle=\frac{1}{\sqrt{2}}(|11'\rangle+|22'\rangle)$. Thus,
by Lemma 2.1, $W$ is an entanglement witness.

If $\rho$ is separable, then ${\rm Tr}((U\otimes
V)W(U^\dagger\otimes V^\dagger)\rho)\geq 0$ as $(U^\dagger\otimes
V^\dagger)\rho(U\otimes V)$ are separable. Conversely, assume that
$\rho=|\psi\rangle\langle\psi|$ is inseparable. Consider its Schmidt
decomposition
$|\psi\rangle=\sum_{k=1}^{N_{\psi}}\delta_k|k,k^\prime\rangle$,
where $\delta_1\geq\delta_2\geq \cdots
> 0$ with $\sum_{k=1}^{N_\psi}\delta_k^2=1$, $\{|k\rangle\}_{k=1}^{N_\psi}$ and
$\{|k^\prime\rangle\}_{k=1}^{N_\psi}$ are orthonormal in $H$ and
$K$, respectively. As $|\psi\rangle$ is inseparable, we must have
its Schmidt number $N_\psi\geq 2$. Thus $\rho=\sum_{k,l
=1}^{N_\psi}\delta_k\delta_{l}|k,k^\prime\rangle\langle l,l^\prime|
$. Up to unitary equivalence, we may assume that
$\{|k\rangle\}_{k=1}^2=\{|i\rangle\}_{i=1}^2$ and
$\{|k^\prime\rangle\}_{k^\prime=1}^2=\{|j'\rangle\}_{j=1}^2$. Then
${\rm Tr}(W\rho)={\rm
Tr}(-\delta_1\delta_2|11'\rangle\langle11'|-\delta_1\delta_2
|22'\rangle\langle22'|)=-2\delta_1\delta_2<0$. Hence the statement
(1) is true.

For the statement (2), assume that there exist $U\in{\mathcal U}(H)$
and $V\in{\mathcal U}(K)$ such that ${\rm Tr}((U\otimes
V)W(U^\dagger\otimes V^\dagger)\rho)< 0$. Then $\rho$ is entangled.
Moreover, $\rho$ has a matrix representation
$$\rho=\sum_{i,j,k,l}\alpha_{ijkl}|Ui\rangle|
Vj^\prime\rangle\langle Uk|\langle Vl^\prime|.$$ Thus, one gets
$$\begin{array}{rl}0>&{\rm Tr}((U\otimes
V)W(U^\dagger\otimes V^\dagger)\rho)={\rm Tr}(W(U^\dagger\otimes
V^\dagger)\rho(U\otimes V))\\
=&{\rm Tr}(\sum_{i,j,k,l}\alpha_{ijkl}(|1\rangle|2'\rangle\langle
1|\langle2'|-|1\rangle|1'\rangle\langle 2|\langle2'|
-|2\rangle|2'\rangle\langle 1|\langle1'|+|2\rangle|1'\rangle\langle
2|\langle1'|)\\
&\cdot(U^\dagger\otimes V^\dagger)|Ui\rangle|
Vj^\prime\rangle\langle Uk|\langle Vl^\prime|(U\otimes V))\\
=&{\rm Tr}(\sum_{i,j,k,l}\alpha_{ijkl}(|1\rangle|2'\rangle\langle
1|\langle2'|-|1\rangle|1'\rangle\langle 2|\langle2'|
-|2\rangle|2'\rangle\langle 1|\langle1'|+|2\rangle|1'\rangle\langle
2|\langle1'|)\\
&\cdot|i\rangle|
j^\prime\rangle\langle k|\langle l^\prime|)\\
=&-\alpha_{2211}-\alpha_{1122}.
\end{array}$$
Now let  $P$ and $Q$ be the projectors from $H$ and $K$ onto the two
dimensional subspaces spanned by $\{|1\rangle,|2\rangle\}$ and
$\{|1^\prime\rangle,|2^\prime\rangle\}$, respectively. Then $${\rm
Tr}(P\otimes Q)(U\otimes V)W(U^\dagger\otimes V^\dagger)(P\otimes
Q)\rho(P\otimes Q))=-\alpha_{2211}-\alpha_{1122}<0,$$ which implies
that $(P\otimes Q)\rho(P\otimes Q)$ is entangled. It follows from
\cite{Hor2} that $\rho$ is 1-distillable. The proof is complete.
\hfill$\Box$

\section{Detecting entanglement of  states by their entries}

In this section, we give a method of detecting entanglement of a
state  in any bipartite system  only by some entries of its matrix
representation.

Let $H$ and $K$ be complex Hilbert spaces of any dimension with
$\{|i\rangle\}_{i=1}^{\dim H}$ and $\{|j'\rangle\}_{j=1}^{\dim K}$
be orthonormal bases of them respectively. Denote
$E_{ij}=E_{i,j}=|i\rangle\langle j|$, which is an operator from
$H$ into $H$. Let  $n\leq\min\{\dim H,\dim K\}$ be a positive
integer. By \cite[Remark 5.2]{QH}, for any permutation $\kappa$ of
$(1,2,\cdots ,n)$, the linear map $\Phi_\kappa:{\mathcal
B}(H)\rightarrow{\mathcal B}(H)$ defined by
$$\Phi_\kappa(A)=(n-1)\sum_{i=1}^n E_{ii}AE_{ii}^\dag +\sum_{i=1}^n
E_{i,\kappa(i)}AE_{i,\kappa(i)}^\dag
-(\sum_{i=1}^nE_{ii})A(\sum_{i=1}^nE_{ii})^\dag \eqno(3.1)$$ for
every $A\in{\mathcal B}(H)$ is a positive elementary operator that
is not completely positive if $\kappa\not= {\rm id}$. Then, for
any unitary operators $U$  and $V$ on $H$, the map
$\Phi_\kappa^{U,V}$ defined by
$$\begin{array}{rl}\Phi_\kappa^{U,V}(A)=&(n-1)\sum_{i=1}^n (VE_{ii}U)A(VE_{ii}U)^\dag +\sum_{i=1}^n
(VE_{i,\kappa(i)}U)A(VE_{i,\kappa(i)}U)^\dag
\\&-(\sum_{i=1}^nVE_{ii}U)A(\sum_{i=1}^nVE_{ii}U)^\dag\end{array}\eqno(3.2)$$
for every $A\in{\mathcal B}(H)$ is positive, too. Let
$\rho_+=|\psi_+\rangle\langle\psi_+|$, where
$$|\psi_+\rangle=\frac{1}{\sqrt{n}}(|1\rangle|1'\rangle +|2\rangle|2'\rangle +\cdots
+|n\rangle|n'\rangle).$$ Then, by Lemma 2.1, we get a class of
entanglement witnesses of the form
$$W_\kappa^{U,V}=n(\Phi_\kappa^{U,V}\otimes I )\rho_+=(\Phi_\kappa^{U,V}(E_{ij})).\eqno(3.3)$$
 Note that $W_\kappa^{U,V}$ is  of finite rank
because $\rho_+$ is.

Particularly, for permutations $\pi, \sigma$ of $(1,2,\cdots ,
n)$, if $U$ and $V$ are the unitary operators defined by
$U^\dag|i\rangle=|\pi(i)\rangle$, $V|i\rangle=|\sigma(i)\rangle$
for $i=1,2,\cdots n$ and $U^\dag|i\rangle=|i\rangle$,
$V|i\rangle=|i\rangle$ for $i>n$, then we have
$$\begin{array}{rl}\Phi_\kappa^{\pi,\sigma}(A)=&\Phi_\kappa^{U,V}(A)=(n-1)\sum_{i=1}^n E_{\sigma(i),\pi(i)}AE_{\sigma(i),\pi(i)}^\dag \\
&+\sum_{i=1}^n
E_{\sigma(i),\pi(\kappa(i))}AE_{\sigma(i),\pi(\kappa(i))}^\dag
-(\sum_{i=1}^n E_{\sigma(i),\pi(i)} )A (\sum_{i=1}^n
E_{\sigma(i),\pi(i)} )^\dag \end{array}\eqno(3.4)$$ for every $A$.
And correspondingly, we get entanglement witnesses of the concrete
form
$$W_\kappa^{\pi,\sigma}=(\Phi_\kappa^{\pi,\sigma}(E_{ij})), \eqno(3.5)$$
where $$\Phi_\kappa^{\pi,\sigma}(E_{ij})=-E_{\sigma(\pi^{-1}(i)),
\sigma(\pi^{-1}(j))} \eqno(3.6)$$ if $1\leq i\not=j\leq n$,
$$\Phi_\kappa^{\pi,\sigma}(E_{ii})=(n-2)E_{\sigma(\pi^{-1}(i)),
\sigma(\pi^{-1}(i))}+E_{\sigma(\kappa^{-1}\pi^{-1}(i)),
\sigma(\kappa^{-1}\pi^{-1}(i))}\eqno(3.7)$$ if $1\leq i\leq n$,
and
$$\Phi_\kappa^{\pi,\sigma}(E_{ij})=0 \eqno(3.8)$$
if $i>n$ or $j>n$.

Thus we have proved the following result.

{\bf Theorem 3.1.} {\it Let $H$ and $K$ be complex Hilbert spaces
of any dimension with $\{|i\rangle\}_{i=1}^{\dim H\leq\infty}$ and
$\{|j'\rangle\}_{j=1}^{\dim K\leq \infty}$ be orthonormal bases of
them respectively. For any positive integer $2\leq n\leq
\min\{\dim H,\dim K\}$ and any permutations $\kappa,\pi,\sigma$ of
$(1,2,\cdots, n)$ with $\kappa\not={\rm id}$, the finite rank
operator $W_\kappa^{\pi,\sigma}$ defined by
$$\begin{array}{rl} W_\kappa^{\pi,\sigma}= &(n-2)\sum_{i=1}^n
|\sigma\pi^{-1}(i),i'\rangle\langle\sigma\pi^{-1}(i),i'|\\
& +\sum_{i=1}^n
|\sigma\kappa^{-1}\pi^{-1}(i),i'\rangle\langle\sigma\kappa^{-1}\pi^{-1}(i),i'|\\
&-\sum_{1\leq i\not=j\leq n}
|\sigma\pi^{-1}(i),i'\rangle\langle\sigma\pi^{-1}(j),j'|
\end{array}$$
is an entanglement witness.}

Assume that $\dim H=\dim K=n$. By applying the witnesses
$W_\kappa^{\pi,\sigma}$ in Theorem 3.1, we get a method of detecting
the entanglement of states by the entries of their density matrix.
Write the product basis of $H\otimes K$ in the order
$$\begin{array}{r}\{|e_1\rangle=|1\rangle|1'\rangle,|e_2\rangle
=|2\rangle|1'\rangle,\cdots,|e_n\rangle=|n\rangle|1'\rangle,
|e_{n+1}\rangle=|1\rangle|2'\rangle,\\
\cdots,|e_{n^2-1}\rangle=|(n-1)\rangle|n'\rangle,|e_{n^2}\rangle
=|n\rangle|n'\rangle\}.\end{array}\eqno(3.9)$$ Then every state
$\rho\in{\mathcal S}(H\otimes K)$ has a matrix representation
$\rho=(\alpha_{kl})_{n^2\times n^2}$.

{\bf Theorem 3.2.} {\it Let $\rho\in{\mathcal B}(H\otimes K)$ with
$\dim H=\dim K=n<\infty$ be a state with the matrix representation
$\rho=(\alpha_{kl})_{n^2\times n^2}$ with respect to the product
basis in Eq.(3.9). If there exist distinguished positive integers
$(i-1)n<k_i, h_i\leq in$, $i=1,2,\cdots,n$
 such that $$\sum_{i=1}^n
k_i=\sum_{i=1}^{n} h_i=\frac{1}{2}n(n^2+1),\eqno(3.10)$$ and
$$(n-2)\sum_{i=1}^n\alpha_{k_ik_i}+\sum_{i=1}^{n}\alpha_{h_ih_i}-\sum_{1\leq
i\not=j\leq n}\alpha_{k_ik_j}<0,\eqno(3.11)$$then $\rho$ is
entangled.}

{\bf Proof.} Eq.(3.10) implies that, there exist permutations
$\pi_1$ and $\sigma_1$ such that $(k_1,k_2-n,\cdots
,k_n-(n-1)n)=\pi_1 (1,2,\cdots, n)$ and $(h_1,h_2-n,\cdots
,h_n-(n-1)n)=\sigma_1 (1,2,\cdots, n)$. It is clear that
$\pi_1(i)\not=\sigma_1(i)$ as $k_i\not=h_i$ for every $i=1,2,\cdots,
n$.

For any permutations $\kappa$, $\pi$ and $\sigma$, by Theorem 3.1,
we have
$$\begin{array}{rl}{\rm
Tr}(W_\kappa^{\pi,\sigma}\rho)=&(n-2)\sum_{i=1}^n\alpha_{\sigma(\pi^{-1}(i))+(i-1)n,\sigma(\pi^{-1}(i))+(i-1)n}\\
&+\sum_{i=1}^n\alpha_{\sigma(\kappa^{-1}\pi^{-1}(i))+(i-1)n,\sigma(\kappa^{-1}\pi^{-1}(i))+(i-1)n}\\&
-\sum_{i\not=j}^n\alpha_{\sigma(\pi^{-1}(i))+(i-1)n,\sigma(\pi^{-1}(j))+(j-1)n}.\end{array}\eqno(3.12)
$$
Comparing Eq.(3.11) with Eq.(3.12), we have to find permutations
$\kappa$, $\pi$ and $\sigma$ so that
$$ \pi_1(i)=\sigma(\pi^{-1}(i))\quad\mbox{and}\quad
\sigma_1(i)=\sigma(\kappa^{-1}\pi^{-1}(i)) \eqno(3.13)
$$
for each $i$, that is, $\pi_1=\sigma\pi^{-1}$ and
$\sigma_1=\sigma\kappa^{-1}\pi^{-1}$. Take $\pi={\rm id}$. Then we
get $\sigma=\pi_1$ and
$\sigma_1=\sigma\kappa^{-1}=\pi_1\kappa^{-1}$. Thus,
$\kappa=\sigma_1^{-1}\pi_1$, $\pi={\rm id}$ and $\sigma=\pi_1$
satisfy Eq.(3.13). With such $\kappa, \pi$ and $\sigma$, by
Eqs.(3.11) and (3.12), we have
$${\rm Tr}(W_\kappa^{\pi,\sigma}\rho)=(n-2)\sum_{i=1}^n\alpha_{k_ik_i}+\sum_{i=1}^{n}\alpha_{h_ih_i}-\sum_{1\leq
i\not=j\leq n}\alpha_{k_ik_j}<0.$$ Hence, $\rho$ is entangled with
$W_\kappa^{\pi,\sigma} $ an entanglement witness for it.
\hfill$\Box$

The general version of Theorem 3.2 is the following result, which is
applicable for bipartite systems of any dimension.

{\bf Theorem 3.3.} {\it Let $H$ and $K$ be complex Hilbert spaces
with $\{|i\rangle\}_{i=1}^{\dim H\leq \infty}$ and
$\{|j'\rangle\}_{j=1}^{\dim K\leq \infty}$ be orthonormal bases of
them respectively. Assume that $\rho$ is a state on $H\otimes K$
and $n\leq\min\{\dim H,\dim K\}$ is a positive integer. If there
exist permutations $\pi$ and $\sigma$ of $(1,2,\cdots ,n)$ with
$\pi(i)\not=\sigma(i)$ for any $i=1,2,\cdots , n$ such that
$$(n-2)\sum_{i=1}^n\langle \pi(i),i'|\rho|\pi(i),i'\rangle+\sum_{i=1}^n\langle \sigma(i),i'|\rho|\sigma(i),i'\rangle
-\sum_{1\leq i\not=j\leq n} \langle
\pi(i),i'|\rho|\pi(j),j'\rangle<0,\eqno(3.14)$$ then $\rho$ is
entangled.}

The idea of the proof of Theorem 3.3 is the same as that of Theorem
3.2 and we omit it here.

Theorems 3.2 and 3.3 tell us, some times we can detect the
entanglement of a state by suitably chosen  $n^2+n$ entries of its
matrix representation with respect to some product basis, where
$n\leq\min\{\dim H,\dim K\}$.

To illustrate how to use Theorem 3.2 and Theorem 3.3 to detect
entanglement of a state, we give some examples.

{\bf Example 3.4.} Let $q_1,q_2,q_3 $ be nonnegative numbers with
$q_1+q_2+q_3=1$ and let $a,b,c\in{\mathbb C}$ with $|a|^2\leq
q_2q_3$, $|b|^2\leq q_2q_3$, $|c|^2\leq q_2q_3$. Let $\rho$ be a
state of $3\times 3$ system with matrix representation
$$\rho=\frac{1}{3}\left(\begin{array}{ccccccccc}q_1&0&0&0&q_1&0&0&0&q_1\\
0&q_3&a&0&0&0&0&0&0\\
0&\bar{a}&q_2&0&0&0&0&0&0\\
0&0&0&q_2&0&b&0&0&0\\
q_1&0&0&0&q_1&0&0&0&q_1\\
0&0&0&\bar{b}&0&q_3&0&0&0\\
0&0&0&0&0&0&q_3&c&0\\
0&0&0&0&0&0&\bar{c}&q_2&0\\
q_1&0&0&0&q_1&0&0&0&q_1
\end{array}\right).\eqno(3.15)$$
Note that, $\rho$ in Eq.(3.15) is a new kind of states, and $\rho$
degenerates to the state as that in  \cite[Example 3.3]{QH} when
$a=b=c=0$.

We claim that, if $q_2<q_1$ or $q_3<q_1$, then $\rho$ is entangled.

In fact, choosing $(k_1,k_2,k_3)=(1,5,9)$, $(h_1,h_2,h_3)=(3,4,8)$
or $(2,6,7)$, we have
$$ \sum _{i=1}^3 \alpha_{k_ik_i}+\sum _{i=1}^3 \alpha_{h_ih_i}-\sum _{1\leq i\not=j\leq 3}
\alpha_{k_ik_j}=\frac{1}{3}(3q_1+3q_2-6q_1)=q_2-q_1
$$
or
$$ \sum _{i=1}^3 \alpha_{k_ik_i}+\sum _{i=1}^3 \alpha_{h_ih_i}-\sum _{1\leq i\not=j\leq 3}
\alpha_{k_ik_j}=\frac{1}{3}(3q_1+3q_3-6q_1)=q_3-q_1.$$
 By Theorem 3.2, we see that $\rho$ is entangled if $q_2<q_1$ or
 $q_3<q_1$.

 It is clear that the partial transpose of $\rho$ in Eq.(3.15) with respect to the first subsystem is
$$\rho^{T_1}=\frac{1}{3}\left(\begin{array}{ccccccccc}q_1&0&0&0&0&0&0&0&0\\
0&q_3&\bar{a}&q_1&0&0&0&0&0\\
0&a&q_2&0&0&0&q_1&0&0\\
0&q_1&0&q_2&0&\bar{b}&0&0&0\\
0&0&0&0&q_1&0&0&0&0\\
0&0&0&b&0&q_3&0&q_1&0\\
0&0&q_1&0&0&0&q_3&\bar{c}&0\\
0&0&0&0&0&q_1&c&q_2&0\\
0&0&0&0&0&0&0&0&q_1
\end{array}\right).$$
Particularly, if we take $q_1=\frac{1}{5}$, $q_2=\frac{1}{10}$,
 $q_3=\frac{7}{10}$ and $a=b=c=\frac{1}{20}$, then, by what proved above, we see that $\rho$ is PPT
 entangled
 because its partial transpose has eigenvalues
 $$\{\frac{1}{60}(8\pm\sqrt{61}),\frac{1}{4},\frac{1}{4},\frac{1}{60},\frac{1}{60},\frac{1}{15},\frac{1}{15},\frac{1}{15}\}
 $$
that are all positive.

{\bf Example 3.5.} Let $\rho$ be a state in $4\times 4$ systems with
the matrix
$$\rho=
\frac{1}{4}\left(\begin{array}{cccccccccccccccc}q_1&0&0&0&0&q_1&0&0&0&0&q_1&0&0&0&0&q_1\\
0&q_4&a&0&0&0&0&0&0&0&0&0&0&0&0&0\\
0&\bar{a}&q_3&0&0&0&0&0&0&0&0&0&0&0&0&0\\
0&0&0&q_2&q_2&0&0&0&0&q_2&0&0&0&0&q_2&0\\
0&0&0&q_2&q_2&0&0&0&0&q_2&0&0&0&0&q_2&0\\
q_1&0&0&0&0&q_1&0&0&0&0&q_1&0&0&0&0&q_1\\
0&0&0&0&0&0&q_4&b&0&0&0&0&0&0&0&0\\
0&0&0&0&0&0&\bar{b}&q_3&0&0&0&0&0&0&0&0\\
0&0&0&0&0&0&0&0&q_3&0&0&c&0&0&0&0\\
0&0&0&q_2&q_2&0&0&0&0&q_2&0&0&0&0&q_2&0\\
q_1&0&0&0&0&q_1&0&0&0&0&q_1&0&0&0&0&q_1\\
0&0&0&0&0&0&0&0&\bar{c}&0&0&q_4&0&0&0&0\\
0&0&0&0&0&0&0&0&0&0&0&0&q_4&d&0&0\\
0&0&0&0&0&0&0&0&0&0&0&0&\bar{d}&q_3&0&0\\
0&0&0&q_2&q_2&0&0&0&0&q_2&0&0&0&0&q_2&0\\
q_1&0&0&0&0&q_1&0&0&0&0&q_1&0&0&0&0&q_1
\end{array}\right), \eqno(3.16)$$
where $q_i\geq 0$ with $\sum_{i=1}^4q_i=1$, $|a|^2$, $|b|^2$,
$|c|^2$ and $|d|^2$ are all $\leq q_3q_4$. $\rho$ defined by
Eq.(3.16) is also a new example, and when $a=b=c=d=0$ we get states
in \cite[Example 4.4]{QH}.

We claim that, if $q_i<q_1$ for some $i\in\{2,3,4\}$; or if
$q_i<q_2$ for some $i\in\{1,3,4\}$, then $\rho$ is entangled.

In fact, we can take
$$(k_1,k_2,k_3,k_4)=(1,6,11,16)\quad\mbox{and}\quad
(h_1,h_2,h_3,h_4)=(2,7,12,13),
$$
or
$$(k_1,k_2,k_3,k_4)=(1,6,11,16)\quad\mbox{and}\quad
(h_1,h_2,h_3,h_4)=(3,8,9,14),
$$
or
$$(k_1,k_2,k_3,k_4)=(1,6,11,16)\quad\mbox{and}\quad
(h_1,h_2,h_3,h_4)=(4,5,10,15),
$$
or
$$(k_1,k_2,k_3,k_4)=(4,5,10,15)\quad\mbox{and}\quad
(h_1,h_2,h_3,h_4)=(1,6,11,16),
$$or
$$(k_1,k_2,k_3,k_4)=(4,5,10,15)\quad\mbox{and}\quad
(h_1,h_2,h_3,h_4)=(2,7,12,13),
$$or
$$(k_1,k_2,k_3,k_4)=(4,5,10,15)\quad\mbox{and}\quad
(h_1,h_2,h_3,h_4)=(3,8,9,14).
$$
Then, it follows from the first three choices that
$$ 2\sum _{i=1}^4 \alpha_{k_ik_i}+\sum _{i=1}^4 \alpha_{h_ih_i}-\sum _{1\leq i\not=j\leq 3}
\alpha_{k_ik_j}=q_i-q_1
$$
with $i=2,3,4$. Hence, by Theorem 3.2 we see that $\rho$ is
entangled if there exists some $i\in\{2,3,4\}$ such that $q_i<q_1$.
Similarly, by the last three choices one sees that $\rho$ is
entangled if there exists some $i\in\{1,3,4\}$ such that $q_i<q_2$.

The kind of states in Eq.(3.16) allow us give some new examples of
entangled states that can not be recognized by PPT criterion and the
realignment criterion. It is obvious that the partial transpose of
$\rho$ in Eq.(3.16) with respect to the first subsystem is
$$\rho^{T_1}=
\frac{1}{4}\left(\begin{array}{cccccccccccccccc}q_1&0&0&0&0&0&0&q_2&0&0&0&0&0&0&0&0\\
0&q_4&\bar{a}&0&q_1&0&0&0&0&0&0&q_2&0&0&0&0\\
0&a&q_3&0&0&0&0&0&q_1&0&0&0&0&0&0&q_2\\
0&0&0&q_2&0&0&0&0&0&0&0&0&q_1&0&0&0\\
0&q_1&0&0&q_2&0&0&0&0&0&0&0&0&0&0&0\\
0&0&0&0&0&q_1&0&0&q_2&0&0&0&0&0&0&0\\
0&0&0&0&0&0&q_4&\bar{b}&0&q_1&0&0&q_2&0&0&0\\
q_2&0&0&0&0&0&b&q_3&0&0&0&0&0&q_1&0&0\\
0&0&q_1&0&0&q_2&0&0&q_3&0&0&\bar{c}&0&0&0&0\\
0&0&0&0&0&0&q_1&0&0&q_2&0&0&0&0&0&0\\
0&0&0&0&0&0&0&0&0&0&q_1&0&0&q_2&0&0\\
0&q_2&0&0&0&0&0&0&c&0&0&q_4&0&0&q_1&0\\
0&0&0&q_1&0&0&q_2&0&0&0&0&0&q_4&\bar{d}&0&0\\
0&0&0&0&0&0&0&q_1&0&0&q_2&0&d&q_3&0&0\\
0&0&0&0&0&0&0&0&0&0&0&q_1&0&0&q_2&0\\
0&0&q_2&0&0&0&0&0&0&0&0&0&0&0&0&q_1
\end{array}\right)$$and that the realignment
 of $\rho$ is
$$\rho^R=
\frac{1}{4}\left(\begin{array}{cccccccccccccccc}q_1&0&0&0&0&q_4&\bar{a}&0&0&a&q_3&0&0&0&0&q_2\\
0&q_1&0&0&0&0&0&0&0&0&0&0&q_2&0&0&0\\
0&0&q_1&0&0&0&0&0&0&0&0&0&0&q_2&0&0\\
0&0&0&q_1&0&0&0&0&0&0&0&0&0&0&q_2&0\\
0&0&0&q_2&q_1&0&0&0&0&0&0&0&0&0&0&0\\
q_2&0&0&0&0&q_1&0&0&0&0&q_4&\bar{b}&0&0&b&q_3\\
0&q_2&0&0&0&0&q_1&0&0&0&0&0&0&0&0&0\\
0&0&q_2&0&0&0&0&q_1&0&0&0&0&0&0&0&0\\
0&0&0&0&0&0&0&q_2&q_1&0&0&0&0&0&0&0\\
0&0&0&0&q_2&0&0&0&0&q_1&0&0&0&0&0&0\\
q_3&0&0&\bar{c}&0&q_2&0&0&0&0&q_1&0&c&0&0&q_4\\
0&0&0&0&0&0&q_2&0&0&0&0&q_1&0&0&0&0\\
0&0&0&0&0&0&0&0&0&0&0&q_2&q_1&0&0&0\\
0&0&0&0&0&0&0&0&q_2&0&0&0&0&q_1&0&0\\
0&0&0&0&0&0&0&0&0&q_2&0&0&0&0&q_1&0\\
q_4&\bar{d}&0&0&d&q_3&0&0&0&0&q_2&0&0&0&0&q_1
\end{array}\right).$$
If we take $q_1=\frac{1}{20}$, $q_2=\frac{1}{10}$,
 $q_3=q_4=\frac{17}{40}$ and $a=b=c=d=\frac{1}{40}$,   $\rho$ is PPT
 entangled
 because $q_1<q_2$ and its partial transpose $\rho^{T_1}$ has eigenvalues
 $$\begin{array}{rl}&\{0.0054,0.0054,0.0069,0.0069,0.0223,0.0223,0.0235,0.0235,\\
    &0.0821,0.0821,0.1027,0.1027,0.1212,0.1212,0.1359,0.1359\}\end{array}
 $$
that are all positive.  Moreover, the trace norm of the realignment
$\rho^R$ of $\rho$ is $\|\rho^R\|_1\doteq 0.8303<1$.  Hence,  we get
another example of entangled states that is PPT and cannot be
detected by the realignment criterion.

It is not difficult to give some examples of applying Theorem 3.3 to
infinite dimensional systems based on examples 3.4 and 3.5.

\section{Conclusions}

Let $H$ and $K$ be  Hilbert spaces and let
$\{|i\rangle\}_{i=1}^{\dim H\leq\infty}$ and
$\{|j'\rangle\}_{j=1}^{\dim K\leq\infty}$ be any orthonormal bases
of $H$ and $K$, respectively. By the finite rank elementary
operator criterion \cite{H}, a state $\rho$ on $H\otimes K$ is
entangled if and only if there exists a finite rank positive
elementary operator $\Phi:{\mathcal B}(H)\rightarrow {\mathcal
B}(K)$ that is not completely positive such that $(\Phi\otimes
I)\rho$ is not positive. By this criterion and the finite rank
positive elementary operators constructed in \cite{QH}, we
construct a collection of finite rank entanglement witnesses.

By using these witnesses we obtain a rank-4 entanglement witness
$W=|1\rangle|2'\rangle\langle
1|\langle2'|-|1\rangle|1'\rangle\langle 2|\langle2'|
-|2\rangle|2'\rangle\langle 1|\langle1'|+|2\rangle|1'\rangle\langle
2|\langle1'|$
 which is universal for pure states, that is,
 for a pure state $\rho$,  $\rho$ is separable if and
only if ${\rm Tr}((U\otimes V)W(U^\dagger\otimes V^\dagger)\rho)\geq
0$ holds for all unitary operators $U$ on $H$ and $V $ on $K$. In
addition, for a mixed state $\rho$, if there exist unitary operators
$U_0$ on $H$ and $V_0$ on $K$ such that ${\rm Tr}((U_0\otimes
V_0)W(U_0^\dagger\otimes V_0^\dagger)\rho)< 0$, then $\rho$ is
entangled and 1-distillable.

Another interesting result, maybe for the first time,  gives a way
of detecting the entanglement of a state in $H\otimes K$ by only a
part entries of its density matrix. This method is simple,
computable and practicable. Assume that $\rho$ is a state on
$H\otimes K$ and $n\leq\min\{\dim H,\dim K\}$ is a positive
integer. If there exist permutations $\pi$ and $\sigma$ of
$(1,2,\cdots ,n)$ with $\pi(i)\not=\sigma(i)$ for any
$i=1,2,\cdots , n$ such that
$$(n-2)\sum_{i=1}^n\langle \pi(i),i'|\rho|\pi(i),i'\rangle+\sum_{i=1}^n\langle \sigma(i),i'|\rho|\sigma(i),i'\rangle
-\sum_{1\leq i\not=j\leq n} \langle
\pi(i),i'|\rho|\pi(j),j'\rangle<0,$$ then $\rho$ is entangled.
Thus we provide a way of detecting the entanglement of a state by
finite suitably chosen entries of its matrix representation with
respect to some product basis. As an illustration how to use this
method, some new examples of entangled states that can be
recognized by this way are proposed, which also provides some new
entangled states that can not be detected by the PPT criterion and
the realignment criterion.


\end{document}